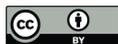

Atmospheric
Measurement
Techniques

Open Access

# Retrieval of nitric oxide in the mesosphere and lower thermosphere from SCIAMACHY limb spectra

S. Bender[1], M. Sinnhuber[1], J. P. Burrows[2], M. Langowski[2], B. Funke[3], and M. López-Puertas[3]

[1]Karlsruhe Institute of Technology, Karlsruhe, Germany
[2]University of Bremen, Bremen, Germany
[3]Instituto de Astrofísica de Andalucía, CSIC, Granada, Spain

*Correspondence to:* S. Bender (stefan.bender@kit.edu)

**Abstract.** We use the ultra-violet (UV) spectra in the range 230–300 nm from the SCanning Imaging Absorption spectroMeter for Atmospheric CHartographY (SCIAMACHY) to retrieve the nitric oxide (NO) number densities from atmospheric emissions in the gamma-bands in the mesosphere and lower thermosphere. Using 3-D ray tracing, a 2-D retrieval grid, and regularisation with respect to altitude and latitude, we retrieve a whole semi-orbit simultaneously for the altitude range from 60 to 160 km.

We present details of the retrieval algorithm, first results, and initial comparisons to data from the Michelson Interferometer for Passive Atmospheric Sounding (MIPAS). Our results agree on average well with MIPAS data and are in line with previously published measurements from other instruments. For the time of available measurements in 2008–2011, we achieve a vertical resolution of 5–10 km in the altitude range 70–140 km and a horizontal resolution of about 9° from 60° S–60° N. With this we have independent measurements of the NO densities in the mesosphere and lower thermosphere with approximately global coverage. This data can be further used to validate climate models or as input for them.

## 1 Introduction

Observations of a number of Antarctic and Arctic winters have shown that in polar regions, downwelling of nitric oxides ($NO_x$ = N, NO, $NO_2$) from the upper mesosphere and lower thermosphere can provide a significant source of $NO_x$ in the polar upper and middle stratosphere, see, for example, Siskind et al. (2000), Funke et al. (2005a), and Randall et al. (2009). However, it is not clear whether the source of these $NO_x$ intrusions is from auroral production in the lower thermosphere, or due to local production in the mesosphere by relativistic electrons, see, e.g. Callis et al. (1998) and Sinnhuber et al. (2011, 2012). The measurements with the SCanning Imaging Absorption spectroMeter for Atmospheric CHartographY (SCIAMACHY) on the European research satellite Envisat quantify this important coupling between the sun and the upper atmosphere as they are the only measurements with a good sensitivity, temporal resolution, and global coverage in the middle and upper mesosphere (70–90 km).

SCIAMACHY was proposed in the 1988 to fly on the then Polar Orbiting Earth Mission, POEM-1 of the European Space Agency, see Burrows et al. (1995) and references therein. SCIAMACHY was selected as a nationally funded contribution: its industrial development being funded by German Aerospace (then DARA and later DLR). Subsequently the Dutch space agency (then NIVR, later NSO) joined the SCIAMACHY agency consortium in Phase A and then Belgium in Phase B. The scientific objectives of SCIAMACHY comprise, beside the study of trace gases in the upper troposphere and stratosphere, the study of the upper atmosphere. Thus the measurement of UV and visible atmospheric emission by molecules and atoms and retrieval of vertical profiles of atmospheric constituents in the mesosphere were targeted. SCIAMACHY has several different measurement modes. The SCIAMACHY proposal envisaged two spectrometers making passive remote-sensing solar and lunar occultation measurements, coupled with limb or nadir measurements of the upwelling radiation at the top of the atmosphere on the sunlit side of the orbit. In phases A (feasibility study) and B (definition study), SCIAMACHY was descoped to one instrument making solar lunar occultation and alternate limb and nadir measurements (Burrows et al., 1995; Bovensmann et al., 1999). As a result of data rate limitations on Envisat, the limb scanning mode of SCIAMACHY was restricted from −2 km to about 93 km. However, in order to probe the mesosphere and thermosphere, the mesosphere and lower thermosphere (MLT) measurements mode was adopted as an additional limb mode since 2008. This manuscript addresses studies using this measurements mode.

We use a part of SCIAMACHY's UV channel 1 (214–334 nm) to infer the NO number densities from atmospheric emissions of the gamma bands in the mesosphere and lower thermosphere. The gamma bands were already used in the retrieval of NO from rocket and satellite experiments. The rocket experiments (Cleary, 1986; Eparvier and Barth, 1992) have a very limited spatio-temporal coverage as they are restricted to single-day flights, in August 1982 (Cleary, 1986) and in March 1989 (Eparvier and Barth, 1992) and to the area close to the starting location, Poker Flat, Alaska, in both cases.

Satellite measurements of nitric oxide (NO) were first done with the SBUV spectrometer on board the Nimbus 7 satellite by Frederick and Serafino (1985) in nadir geometry using the NO (1,4) gamma emission at 255 nm. Further previous satellite experiments using the NO gamma bands in the ultra-violet spectral range include the Student Nitric Oxide Explorer (SNOE, Barth et al., 2003) which measured nitric oxide in the thermosphere (97 to 150 km) from March 1998 until December 2003. Results are published until September 2000 and are used by the NOEM empirical model of nitric oxide in the MLT region (Marsh et al., 2004). The Ionospheric Spectroscopy and Atmospheric Chemistry (ISAAC, Minschwaner et al., 2004) satellite performed measurements at 80–200 km and results are published for November 1999–December 1999.

Measurements of NO in the mesosphere using microwaves were done by the Sub-Millimitre Radiometer (SMR) on board the Odin satellite. Infrared measurements were performed by the Optical Spectrograph and InfraRed Imaging System (OSIRIS) on board the Odin satellite which measured nighttime NO from $NO_2$ airglow emissions up to 100 km (Sheese et al., 2011). Further infrared instruments are the Atmospheric Chemistry Experiment Fourier transform spectrometer (ACE-FTS) (Kerzenmacher et al., 2008) and the Michelson Interferometer for Passive Atmospheric Sounding (MIPAS) also on the Envisat satellite. The MIPAS instrument measures the vibrational–rotational emission lines of NO with its infrared spectrometer directly (Funke et al., 2005b; Bermejo-Pantaleón et al., 2011). It shares the satellite with SCIAMACHY but is oriented to view in the opposite direction. This means that MIPAS samples the air at approximately the same tangent points but with a delay of about 50 min. With its special upper atmosphere (UA) mode which, however, is also not continuous, it provides another independent verification of our results.

This paper is organised as follows: we present some details about SCIAMACHY and its MLT mode in Sect. 2, Sect. 3 goes into detail about the NO gamma bands and the retrieval method. Finally, our results and the comparisons to MIPAS data and the NOEM model are shown in Sect. 4.

## 2 SCIAMACHY MLT mode

SCIAMACHY is a limb sounding experiment aboard the European Envisat satellite flying in a sun-synchronous orbit at approximately 800 km altitude since 2002 (Burrows et al., 1995; Bovensmann et al., 1999). The wavelength range of SCIAMACHY extends from UV to the near infrared (220–2380 nm). From July 2008 to April 2012, SCIAMACHY performed observations in the mesosphere and lower thermosphere (MLT, 50–150 km) regularly twice per month; this limb mesosphere-thermosphere state was coordinated with the MIPAS upper atmosphere (UA) mode once every 30 days.

The MLT mode sampled the mesosphere and lower thermosphere at 30 limb points from 50 to 150 km with a vertical spacing of about 3 km. The scans were scheduled in place of the usual nominal mode limb scans from 0 to 90 km, so that the rest of the measurements within the orbit sequence remained unchanged and there were around 20 limb scans along a semi-orbit. This compares well with the MIPAS measurement sequence, which collects limb scans every four to five degrees and its UA mode scans from 50 to 170 km.

The use of the NO gamma bands for the SCIAMACHY retrieval, where the first electronic state needs to be excited, restricts the useful measurements to times with available sunlight for this excitation. That means we can only retrieve the NO densities at daytime with SCIAMACHY. In contrast to that, the use of the infrared bands by MIPAS allows them to measure the NO densities at night-time as well. We take this into account when comparing the measured densities from both instruments.

## 3 Retrieval algorithm

Fitting the calculated NO emissivities to the calibrated measured spectra gives the slant column densities. These are used to retrieve vertical profiles of number densities using a constrained iterative least squares algorithm based on the work by Rodgers (1976) and which is similar to the one used in the retrieval of the MIPAS data products (von Clarmann et al., 2003; Funke et al., 2005b). Further inputs are the high-resolution solar spectrum (Chance and Kurucz, 2010) and the temperature from the NRLMSISE-00 model (Picone et al., 2002) for the emissivity calculation, as well as the NOEM model (Marsh et al., 2004) as the a priori data for the retrieval.

**Table 1.** NO emissivity parameters used in the paper for the three NO gamma bands used for the retrieval.

|  | (0, 2) | (1, 4) | (1, 5) |
|---|---|---|---|
| $\lambda_{v'0}$ [nm] | 226.5 | 215.1 | 215.1 |
| $\lambda_{v'v''}$ [nm] | 247.4 | 255.4 | 267.4 |
| $f_{v'0}$ [$10^{-4}$] | 3.559 | 7.010 | 7.010 |
| $\omega_{v'v''}$ | 0.2341 | 0.1115 | 0.0971 |

### 3.1 NO gamma bands

The NO gamma bands result from the electronic transition of the NO molecule from the first excited state, $A^2\Sigma^+$, to the ground state $X^2\Pi$. The excitation happens by absorbing solar UV light, which restricts our retrieval to daytime nitric oxide, as discussed at the end of Sect. 2.

The various gamma bands have different advantages and disadvantages for retrieval. The ones to the vibrational ground state of the electronic ground state, e.g. the (1, 0), (2, 0) transitions, have the advantage of high emission rates and the disadvantage of self-absorption, see, for example, Eparvier and Barth (1992) and Stevens (1995). This makes the atmosphere optically thick with respect to these bands and the emissivity depends on the total slant column density along the line-of-sight (Eparvier and Barth, 1992). In contrast to that, transitions to a different vibrational state, e.g. the (0, 1), (0, 2), or (1, 4) transitions, have a lower emissivity, but the probability of self-absorption is greatly reduced.

### 3.2 NO emissivity calculation

The emissivities of the NO gamma bands are calculated following Stevens and others (Eparvier and Barth, 1992; Stevens, 1995). The emission rate factors $g_{v'v''}$ for the transition from the vibrational state $v'$ of the electronically excited state to the vibrational state $v''$ of the electronic ground state are given by the sums over all rotational emission factors $g_{j'j''}$ (see Eparvier and Barth, 1992 and references therein and Stevens et al., 1997), which are given by

$$g_{j'j''} = \frac{S(j'j'')}{2j'+1}\omega_{v'v''} \\ \cdot \sum_{j=|\Lambda+\Sigma|}^{\infty} \frac{\pi e^2}{mc^2}\lambda_{j'j}^2 \pi F_{j'j} f_{v'0} \frac{S(j'j)}{2j+1} \frac{N_j}{N_0}. \quad (1)$$

Here, $S$ are the Hönl–London factors for the rotational transitions (Earls, 1935; Schadee, 1964; Tatum, 1967), $\omega$ is the branching fraction, $\lambda$ the transition wavelength, $\pi F$ is the solar irradiance at that wavelength taken from Chance and Kurucz (2010), and $f$ is the oscillator strength. $N_j$ is the NO population of the rotational level $j$ and $N_0$ the population of the zero vibrational level of the electronic ground state (Eparvier and Barth, 1992). The most important parameters for the emissivity calculation are summarised in Table 1, listing the selected values from Luque and Crosley (1999). The molecular constants can be found in Table 2 of Eparvier and Barth (1992).

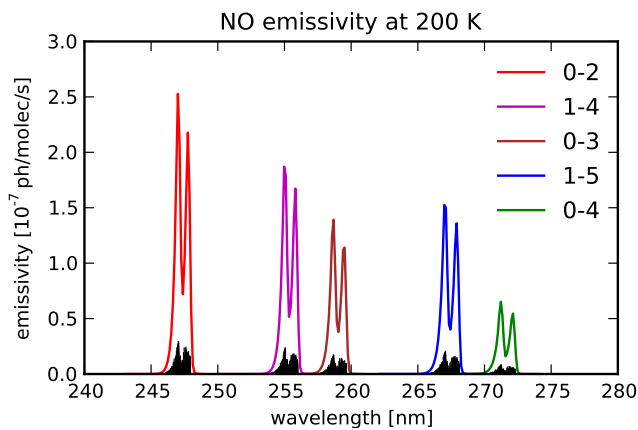

**Fig. 1.** NO emissivities calculated for a temperature of 200 K for a selected number of gamma bands. The small black lines indicate the individual rotational emission lines within the vibrational transition bands and the coloured lines are the emissivity bands resulting from binning to the SCIAMACHY spectral resolution and convolving with the respective spectral slit function.

The calculated spectrum at 200 K for a selected number of gamma bands is shown in Fig. 1, where the small black lines indicate the individual rotational transitions and the larger envelope lines result from binning to the SCIAMACHY resolution and the convolution with the spectral slit function of the instrument. The spectra are calculated using spectroscopic data (oscillator strengths and branching fractions) from Luque and Crosley (1999).

The gamma bands shown in Fig. 1 are the ones that are visible in SCIAMACHY's UV channel 1. For the retrieval we use the three gamma bands with the largest emissivities in that spectral range, which are the (0,2), the (1,4), and the (1,5) bands at 247, 255, and 267 nm, respectively. These are all non-resonant bands and the line-of-sight should be optically thin. Additionally including the (0,3) band gives a negligible improvement of the results but increases the memory needs and processing time because of the larger matrices involved in the retrieval.

The integrated band emission rate factors for 200 and 1000 K are listed in Table 2, with the top line listing the numbers used in previous work (Minschwaner et al., 2004) using the parameters from Piper and Cowles (1986); Stevens (1995). The emission rate factors used in this work are listed in the second line, calculated using the experimentally derived parameters from Luque and Crosley (1999). They agree within a few percent and the differences can be attributed to using slightly different oscillator strengths and branching fractions.

**Table 2.** NO band integrated emission rate factors (in $10^{-6}\,\mathrm{ph\,s^{-1}}$) at 200 and 1000 K for the $\gamma$ bands used in the retrieval. Top row: results from Stevens (1995), using the parameters from Piper and Cowles (1986), and used for NO retrieval by Minschwaner et al. (2004). Bottom row: factors used in this work, using the parameters from Luque and Crosley (1999).

| Reference | $T$ [K] | (0, 2) | (1, 4) | (1, 5) |
|---|---|---|---|---|
| Minschwaner et al. (2004) | 200 | 2.08 | 1.56 | 1.49 |
| cf. Stevens (1995) | 1000 | 2.17 | 1.53 | 1.45 |
| this work | 200 | 2.02 | 1.60 | 1.39 |
|  | 1000 | 2.11 | 1.57 | 1.37 |

The variation of the emissivity over the temperature range from 200 to 1000 K is smaller than five percent, see Table 2. The temperature range found in the thermosphere is smaller (200 K...600 K) such that the error introduced by using model temperatures from NRLMSISE-00 (Picone et al., 2002) instead of temperatures measured by MIPAS (Bermejo-Pantaleón et al., 2011) is small, although deviations around thirty up to sixty percent from the model were found.

### 3.3 Radiative transfer

The general forward model is the functional relation

$$y = F(x) \quad (2)$$

between the measurements $y$ and the quantities of interest $x$. In the case discussed in this paper, the measurements are the spectral irradiances at the frequency $\nu$ (or at wavelength $\lambda$) at the satellite point, $I_\nu$. Of interest are the number densities of NO at the retrieval point $s$, $x(s)$.

The measured irradiance $I_\nu$ at the frequency $\nu$ is given by the radiative transfer equation as the integral over the line-of-sight:

$$I_\nu = \int_{\mathrm{LOS}} (x(s)\gamma_\nu(s)F_{\nu'}(s) + \sigma_R \varrho_{\mathrm{air}})\,\mathrm{e}^{-\tau_\nu}\mathrm{d}s \quad (3)$$

$$\approx \gamma F \mathrm{e}^{-\tau} \int_{\mathrm{LOS}} x(s)\mathrm{d}s = \gamma F \mathrm{e}^{-\tau} \cdot \varrho_{\mathrm{sc}}, \quad (4)$$

In Eq. (3), $x$ is the number density of the species, $\gamma$ the emissivity factor, and $F$ the solar irradiance at the excitation frequency $\nu'$. The Rayleigh part of the spectrum, $\sigma_R \varrho_{\mathrm{air}}$, is fitted as a spectral background signal.

Since we use non-resonant transitions, the line-of-sight is optically thin with respect to self-absorption (Stevens et al., 1997). Retrieving number densities above 60 km has the additional advantage that quenching is negligible and the optical depth is mainly determined by the ozone and air absorption along the line-of-sight and along the line from the sun (Stevens et al., 1997). This absorption is independent of the NO number density $x$ and modifies the column emission rate as seen by the instrument by the constant factor $\mathrm{e}^{-\tau}$. The optical depth $\tau$ is given by

$$\tau = \sum_i \int_{\mathrm{LOS}} \sigma_i^{\mathrm{abs}} \varrho_i^{\mathrm{abs}} \mathrm{d}s, \quad (5)$$

where the sum includes all absorbing species along the line-of-sight. This leaves us with the linear relationship of the intensity to the slant column density $\varrho_{\mathrm{sc}}$ in Eq. (4).

### 3.4 Retrieval algorithm

The general problem is to invert Eq. (2) to extract $x$ from $y$, where there are usually a different number of measurements ($y$) than unknowns ($x$), which means that Eq. (2) is underdetermined or overdetermined. In the case of an overdetermined system, this inversion problem can be solved by minimising

$$\|\mathbf{K}x - y\|^2, \quad (6)$$

where $\mathbf{K}$ is the Jacobian of the forward model Eq. (2). The measurements $y$ are the slant column densities from fitting the calculated NO spectra to the measured ones. We fit each NO gamma band individually to get three separate measurements (and errors) of the slant column density at the tangent points. The measurement vector $y$ has the dimension $3\times$(number of tangent points) and the results from all three gamma bands are used simultaneously to retrieve the number density $x$.

In the case of an underdetermined system, additional constraints are needed. These are given by the a priori input, i.e. approximate knowledge of the solution $x$. The minimisation problem then changes to

$$\|\mathbf{K}x - y\|^2_{\mathbf{S}_y^{-1}} + \|x - x_a\|^2_{\mathbf{S}_a^{-1}}, \quad (7)$$

where $\mathbf{S}$ are the covariance matrices and the subscript "a" indicates the a priori quantities. Doing a 2-D-retrieval, we additionally constrain the solutions to not vary too much with altitude and latitude by introducing regularisation matrices $\mathbf{R}_{\mathrm{alt}}$ and $\mathbf{R}_{\mathrm{lat}}$, see Scharringhausen et al. (2008a) and Langowski et al. (2013). This then leads to the minimisation of

$$\|\mathbf{K}x - y\|^2_{\mathbf{S}_y^{-1}} + \|x - x_a\|^2_{\mathbf{S}_a^{-1}}$$
$$+ \lambda_{\mathrm{alt}} \|\mathbf{R}_{\mathrm{alt}}(x - x_a)\|^2 + \lambda_{\mathrm{lat}} \|\mathbf{R}_{\mathrm{lat}}(x - x_a)\|^2 \quad (8)$$

in our case. The a priori covariance matrix $\mathbf{S}_a$ is chosen as $\lambda_a \mathbf{I}$ with an empirically tuned parameter $\lambda_a$. The regularisation matrices $\mathbf{R}_{\mathrm{alt}}$ and $\mathbf{R}_{\mathrm{lat}}$ are discrete first derivative matrices with respect to altitude and latitude. Their effect is a smoothing of the solution in these directions, preventing large oscillations by carefully choosing $\lambda_{\mathrm{alt}}$ and $\lambda_{\mathrm{lat}}$. The regularisation parameters used in this work are given in Table 3.

**Table 3.** Regularisation parameters as used in the NO retrieval from SCIAMACHY limb scans.

|  | 1-D | 2-D |
| --- | --- | --- |
| $\lambda_{\text{a}}$ | $3 \times 10^{-18}$ | $3 \times 10^{-18}$ |
| $\lambda_{\text{alt}}$ | $1 \times 10^{-17}$ | $1 \times 10^{-17}$ |
| $\lambda_{\text{lat}}$ | 0 | $3 \times 10^{-17}$ |

The solution of Eq. (8) is obtained by an iterative algorithm. Defining a combined regularisation matrix $\mathbf{R}$ as

$$\mathbf{R} := \mathbf{S}_{\text{a}}^{-1} + \lambda_{\text{alt}} \mathbf{R}_{\text{alt}}^T \mathbf{R}_{\text{alt}} + \lambda_{\text{lat}} \mathbf{R}_{\text{lat}}^T \mathbf{R}_{\text{lat}} \,, \quad (9)$$

the intermediate solution is given by (von Clarmann et al., 2003; Funke et al., 2005b)

$$\begin{aligned} \boldsymbol{x}_{i+1} = \boldsymbol{x}_i &+ \left(\mathbf{K}^T \mathbf{S}_y^{-1} \mathbf{K} + \mathbf{R}\right)^{-1} \\ &\cdot \left[\mathbf{K}^T \mathbf{S}_y^{-1} (\boldsymbol{y} - \boldsymbol{y}_i(\boldsymbol{x}_i)) + \mathbf{R}(\boldsymbol{x}_{\text{a}} - \boldsymbol{x}_i)\right]. \quad (10) \end{aligned}$$

## 3.5 Model input

Our retrieval uses model input to provide an initial guess for the number density and to improve the convergence of the iteration. The a priori input stems from the NOEM model (Marsh et al., 2004), which is also used by the MIPAS NO retrievals (Funke et al., 2005b). It is an empirical model based on SNOE observations (Barth et al., 2003) and constructed for the altitude range from 100 to 150 km only (Marsh et al., 2004). The extension downwards and upwards is done by using constant number densities independent of altitude. Hence, the use of the NOEM model for altitudes below 100 km is questionable, but we use it because of our direct comparisons to MIPAS data.

Another input necessary for the emissivity calculation of the NO rotational transitions are the atmospheric temperatures. These are calculated using the NRLMSISE-00 model (Picone et al., 2002) which has as main input parameters the geomagnetic $A_{\text{p}}$ index and the solar radio flux $f_{10.7}$. These data were taken from the Space Physics Interactive Data Resource (SPIDR) service (NGDC and NOAA, 2011) of the National Geophysical Data Center (NGDC) of the National Oceanic and Atmospheric Administration (NOAA).

## 4 Results

### 4.1 NO number density

To illustrate the results of our retrieval, we choose one example orbit, 41 467 from 3 February 2010, to show the individual steps. For one selected tangent point, 71° N, 105 km, the spectral fit for the NO (0,2) transition is shown in Fig. 2. This latitude and altitude were chosen such that the NO content in that region is supposed to be non-negligible as indicated by previous measurements from SNOE (Barth et al., 2003). As the figure shows, the UV spectra can be noisy, i.e. the NO signal is near the noise threshold, introducing errors into the fitted slant column densities and reducing the signal-to-noise ratio (SNR). These errors are taken into account by the retrieval algorithm as the covariance matrix $\mathbf{S}_y$, see Eqs. (7) and (8), giving less weight to points with a low SNR.

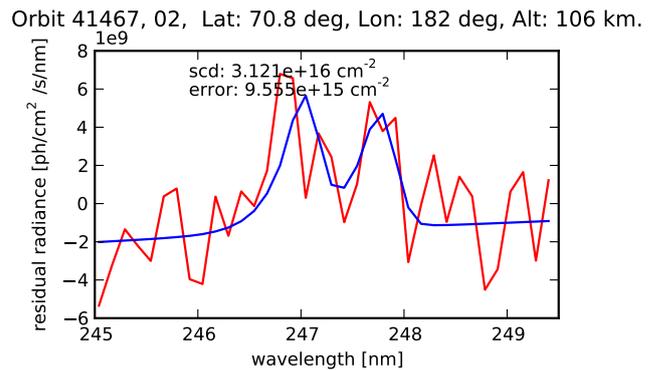

**Fig. 2.** NO (0,2) spectral fit, showing the residual spectra (red) after dark-current subtraction and Rayleigh background fit, and the fitted NO spectrum (blue) with a low order polynomial as a baseline.

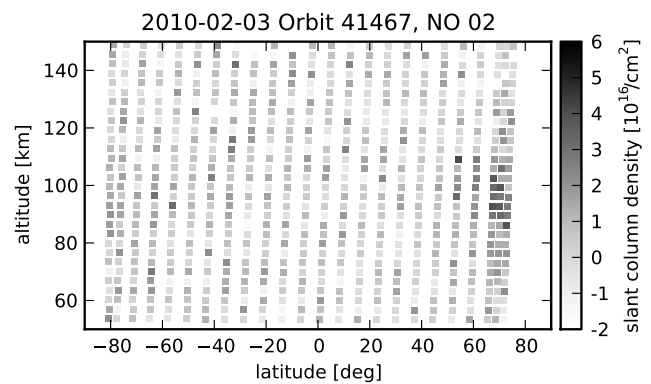

**Fig. 3.** NO slant column densities for the (0,2) transition along the sample orbit (no. 41467, 3 February 2010).

The resulting slant column densities for the same orbit are shown in Fig. 3 for the (0,2) transition. The largest slant column densities for all three lines (the other two are not shown here) are observed at approximately the same tangent points, namely in the region from 60° N–70° N and at altitudes around 100 km.

The zonal mean (60° N–90° N) for the line-of-sight emissivities are shown in the left panel of Fig. 4. They are obtained by integrating the relevant part of the measured spectrum and averaging the results with equal weights. The maximum of the emissivity profile is observed slightly below 100 km limb tangent altitude. The profile of the fitted slant column densities, shown in the middle panel of the same figure, shows the same features and reflects the measured emissivities. The resulting number density profile is shown in the right panel of Fig. 4. It shows that the altitude of the largest number density is above the peak emissivity and peak slant column density. This is the result of the retrieval algorithm that takes into account that the slant column density is given by the total NO emission along the line-of-sight which is largest at a tangent point where the line-of-sight includes the whole NO layer in the thermosphere, i.e. slightly below 100 km.

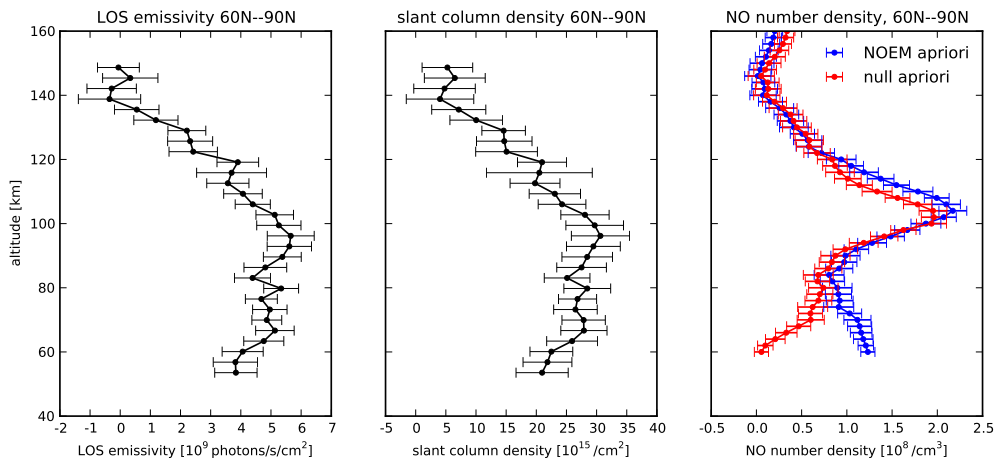

**Fig. 4.** Zonal mean (60° N–90° N) integrated NO emissivity profile (3 February 2010) (left panel), zonal mean fitted slant column densities (middle panel), and zonal mean NO number densities (right panel) using the NOEM model as a priori input (blue line) and without prior assumptions (red line). The error bars show the $3\sigma$ confidence interval.

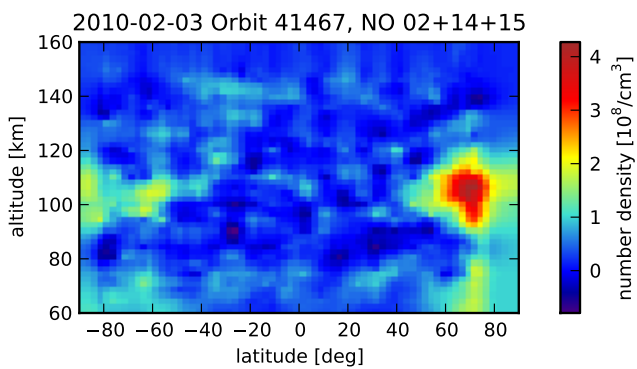

**Fig. 5.** Retrieved NO number densities along the sample orbit (no. 41467, 3 February 2010).

Using the retrieval algorithm as described in Sect. 3, the calculated number density distribution along the orbit is shown in Fig. 5. As already indicated by the slant column densities and the number density profile in Fig. 4, the maximum here is between 60° N and 70° N and at an altitude of around 100–110 km. In the northern winter there is polar night at latitudes north of 80°, hence the retrieval of NO in this region is not possible.

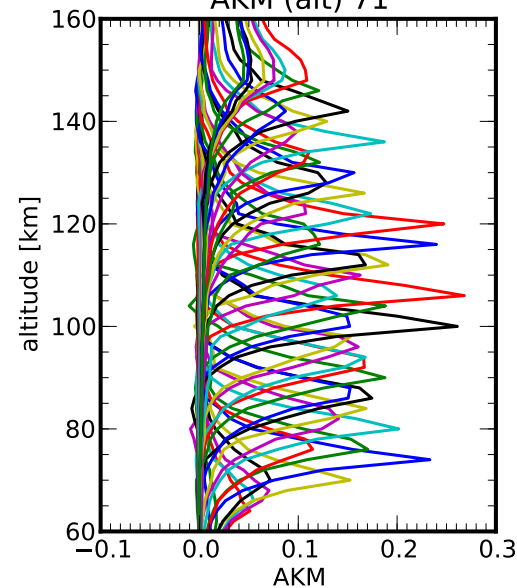

**Fig. 6.** Altitude averaging kernel matrix elements for a sample orbit (no. 41467, 3 February 2010).

### 4.2 Resolution

The vertical averaging kernel elements of our retrieval for a particular orbit are shown in Fig. 6. The respective full width at half maximum, fwhm, are shown in Fig. 7. The fwhm is used as a measure of the vertical resolution and is 10 km or better in the range from 70 to 150 km. A vertical resolution of about 5 km is achieved from 80 to 140 km.

The horizontal averaging kernel elements, calculated for the same orbit are presented in Fig. 8 with the respective fwhm shown in Fig. 9. In the retrieval runs in this paper, the grid consists of 2.5° latitude bins, which restricts the fwhm to the resolution within subsequent 2.5° bins. This makes the horizontal resolution jump between 7.5° and 12.5° around the true value. The average is obtained by a cubic B-spline fit using a large smoothing constraint, and a similar figure is obtained by using a 13 or 15 point running mean. It shows that the real horizontal resolution, determined by the radiative transfer, is quite constant between 60° S and 60° N, being that for 9°.

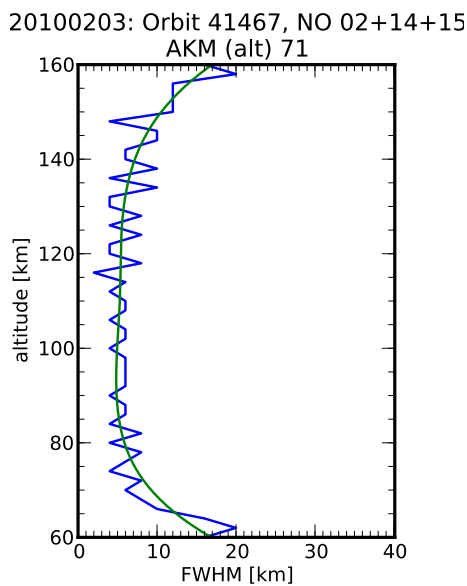

**Fig. 7.** FWHM of the altitude averaging kernel matrix elements for a sample orbit (no. 41467, 3 February 2010) as an indicator of vertical resolution.

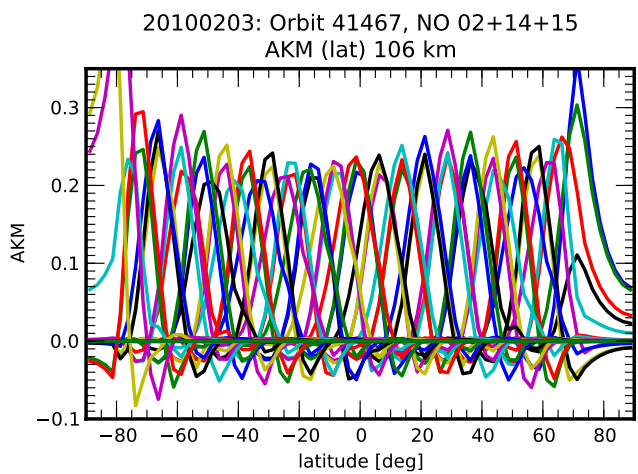

**Fig. 8.** Latitude averaging kernel matrix elements for a sample orbit (no. 41467, 3 February 2010).

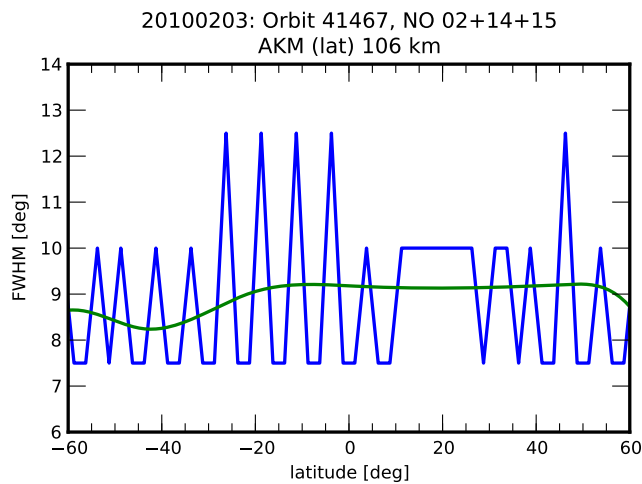

**Fig. 9.** FWHM of the latitude averaging kernel matrix elements for a sample orbit (no. 41467, 3 February 2010). The apparent large variation is due to the grid spacing of 2.5°. The green line is a smoothed cubic B-spline fit indicating an average latitudinal resolution of about 9°.

### 4.3 Comparison to one-dimensional retrieval

A common approach used in trace gas retrieval consists of using homogeneous atmosphere layers to invert only a single limb scan, resulting in a vertical profile at one geolocation. This works well when the radiative paths through the atmospheric volumes for each scan are independent of one another, i.e. the distance between limb scans is larger than the optical path for a given tangent height and the gradient in NO is small. NO in the lower thermosphere and upper mesosphere, however, has a steep horizontal gradient from high to mid-latitudes. Here, the two-dimensional approach allows us to take advantage of the closely spaced limb scans from SCIAMACHY, in particular in the northern polar region. The combined information from the overlapping limb scans along the line-of-sight helps to better represent the strong latitudinal gradient. The method is new, and is described in more detail in Langowski et al. (2013). To demonstrate the advantage of the 2-D, or more generally tomographic, retrieval over the 1-D retrieval, a comparison is presented.

The 1-D version of the retrieval results in a vertical profile of the NO number density along a single limb scan. In the 2-D version of the retrieval, the latitude distribution of a limb scan spans three latitude bins of the retrieval grid. To compare those two methods, the number densities of the 2-D result have been interpolated to the limb scan's latitudes. The vertical profiles in comparison are shown in Fig. 10 together with the a priori input. Results are shown for high northern latitude (61° N), in the equatorial region (10° N), and in the southern polar region (63° S).

The profiles agree within the error bars (measurement errors) but are not identical. It is also clear that the a priori is not the primary input for the retrieved number density since there are both, differences in the peak altitude and in the absolute value of the density. The 2-D results show a smaller measurement error due to the possible influence of neighbouring limb scans in the retrieval. However, to make both methods comparable, the latitudinal regularisation was switched off for the 1-D version and the vertical parameter was slightly adjusted. This comparison gives confidence in the 2-D retrieval since the results are of the same quality or better than the 1-D results. Thus, it is a valid method to measure the NO content in the middle atmosphere from SCIAMACHY UV spectra.

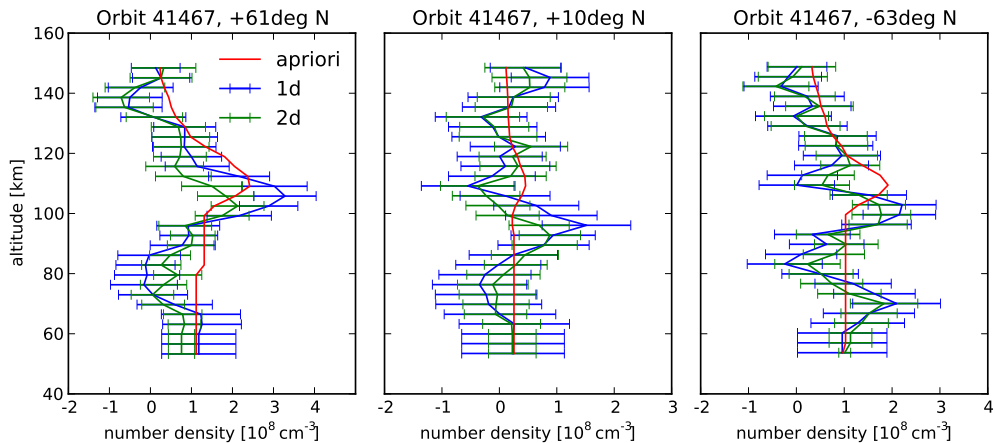

**Fig. 10.** Comparison of retrieval methods, 1-D retrieval (blue line), 2-D retrieval (green line), and the respective a priori input (red line) for three geographic regions of the selected sample orbit (no. 41467, 3 February 2010).

### 4.4 Comparison to MIPAS measurements

As MIPAS and SCIAMACHY both fly on Envisat, although viewing in the opposite direction, a direct comparison of the results is possible. Data products from the MIPAS instrument (Bermejo-Pantaleón et al., 2011) include the volume mixing ratio (vmr) rather than the number densities of NO. We have converted the MIPAS vmr values to number densities using their retrieved temperatures. Retrieved polar NO profiles have a vertical resolution of 5–10 km for high $A_\mathrm{p}$ values, and degrade to 10–20 km for low $A_\mathrm{p}$ conditions. Below approximately 120 km, total retrieval errors are dominated by instrumental noise. The typical single measurement precision of NO abundances is 10–30 % for high $A_\mathrm{p}$ values, increasing to 20–50 % for low $A_\mathrm{p}$ conditions. Above 120 km, systematic errors due to uncertainties of the atomic oxygen profile used in the non-LTE modelling are the dominant error source.

We compare the vertical NO columns in the range of 70–140 km from MIPAS to the SCIAMACHY derived results. This is the range of maximal vertical resolution of the SCIAMACHY retrieval. Both instruments have different vertical resolutions, and both are broad with respect to NO vertical variability. For this reason, column comparisons have been done. This, however, implies merging two different MIPAS NO products (below 100 km using temperature retrieved at 15 μm and above from a joint $T$-NO retrieval at 5.3 μm) which may have different biases. Nevertheless, possible inconsistencies related to this merging are considered to have a smaller impact on the column comparisons than the different vertical resolutions would have on profile comparisons. For our comparison, the vertical column ratios (SCIAMACHY/MIPAS) for orbits, where both instruments scanned the upper atmosphere are binned into 20° latitude bins and are shown for 2008 in Fig. 11 and for 2009 in Fig. 12. The MIPAS measurements are confined to those for daylight limb scans, selected by using the criterion that the solar zenith angle must be smaller than 95°, since the comparison goes down to 70 km.

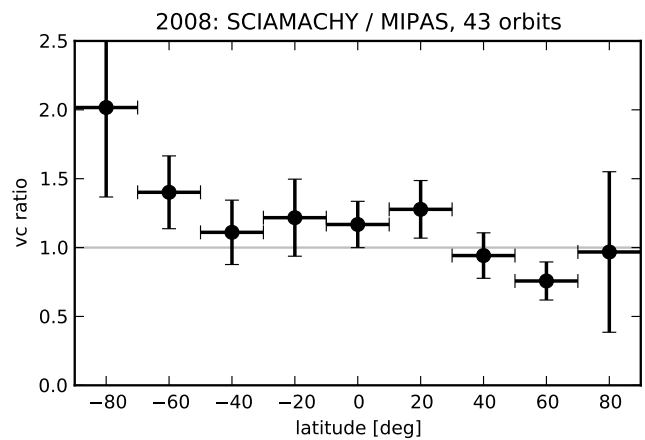

**Fig. 11.** Ratio of SCIAMACHY to MIPAS NO vertical columns (70–140 km), averaged in 20° latitude bins from 43 coincident orbits in 2008. The horizontal bars indicate the bins and the vertical bars the $3\sigma$ confidence interval.

For the majority of latitudes, the ratio of the SCIAMACHY to MIPAS NO columns is close to one considering the standard error of the mean (error bars). We find that the SCIAMACHY measured columns are in general larger than the MIPAS results, except for higher northern latitudes (in particular in 2008). One possible explanation might be the deficit of simultaneous measurements in that region. There are four times (2009) to ten times (2008) as many coincident points at middle to low latitudes as in the 80° bins. This ratio is about two in the −80° bins and although both retrievals use a similar regularisation, different biases can result when dealing with unavailable direct measurements. The a priori constraint in the SCIAMACHY retrieval, $S_a$, bends the solution into the direction of the number density given by the model. In contrast, the MIPAS regularisation matrices have the effect of using the profile shape as a constraint for the so-lution. This may explain part of the disagreement, but further research is required to establish the origin of the differences quantitatively.

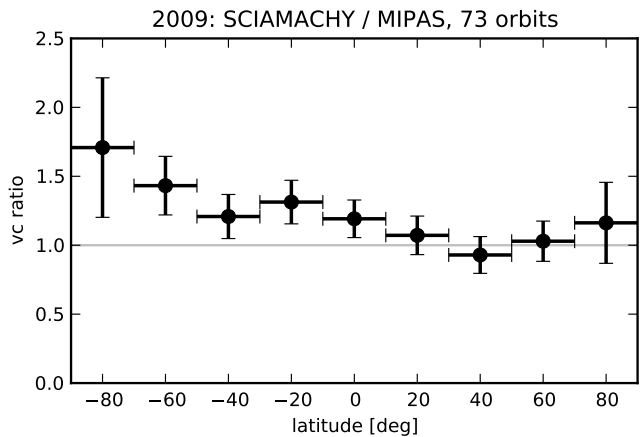

**Fig. 12.** Ratio of SCIAMACHY to MIPAS NO vertical columns (70–140 km), averaged in 20° latitude bins from 73 coincident orbits in 2009. The horizontal bars indicate the bins and the vertical bars the $3\sigma$ confidence interval.

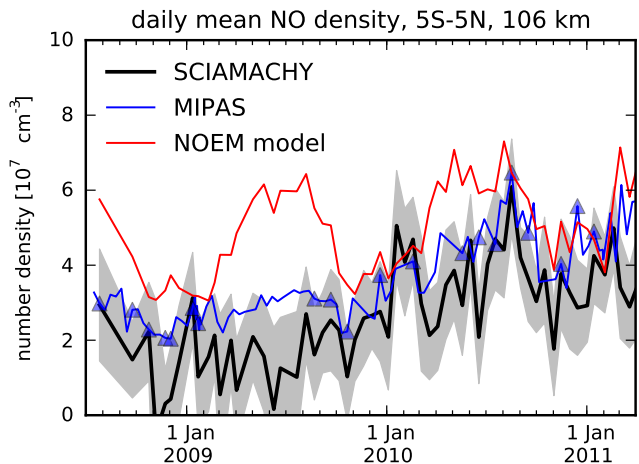

**Fig. 13.** Equatorial daily mean nitric oxide number density at 106 km, SCIAMACHY MLT (black) and MIPAS UA (blue) compared to the NOEM model (red) (Marsh et al., 2004) based on SNOE data (Barth et al., 2003). The grey shaded area shows the $3\sigma$ confidence interval of the SCIAMACHY daily means and the blue triangles indicate coincident SCIAMACHY MLT and MIPAS UA measurement days.

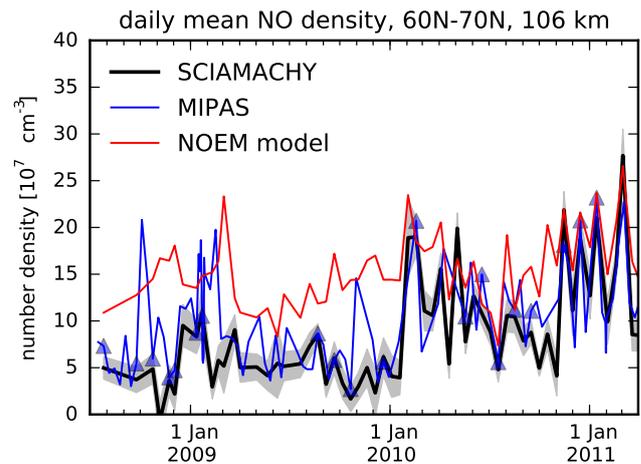

**Fig. 14.** Auroral daily mean nitric oxide number density at 106 km, SCIAMACHY MLT (black) and MIPAS UA (blue) compared to the NOEM model (red) (Marsh et al., 2004) based on SNOE data (Barth et al., 2003). The grey shaded area shows the $3\sigma$ confidence interval of the SCIAMACHY daily means and the blue triangles indicate coincident SCIAMACHY MLT and MIPAS UA measurement days.

### 4.5 Number density time series

For the MLT observations from July 2008 until March 2011, the NO number density at 106 km in the equatorial region (5° S–5° N) is shown in Fig. 13, in the auroral region (60° N–66° N) in Fig. 14. We compare the results for NO, retrieved from SCIAMACHY with those from MIPAS and those from the NOEM model (Marsh et al., 2004) which is derived from SNOE measurements (Barth et al., 2003).

The patterns of the SCIAMACHY NO are similar to those from NOEM. However, the NOEM results are systematically higher than the SCIAMACHY data. This disagreement is largest in 2008 and 2009, during the deep solar minimum, when also the geomagnetic activity was very low; in 2010, when both solar and geomagnetic activity were beginning to increase again, the agreement is better. Nearly perfect agreement is reached in 2011 at high latitudes. The consistency among the MIPAS and SCIAMACHY data may hint at assumptions within the NOEM model, which was derived from SNOE observations during a period of continuous high solar and geomagnetic activity (1998–2000).

## 5 Conclusions

We present a retrieval algorithm for the extraction of NO number densities in the mesosphere and lower thermosphere (MLT) region from the measurements of scattered solar radiation and atmospheric emissions in limb viewing geometry from the SCIAMACHY instrument aboard Envisat. The method is adapted from earlier work (cf. Scharringhausen et al., 2008b, a) and was extended to use the gamma bands of NO and the UV spectra of SCIAMACHY's special MLT limb scans. The retrieval method is an adapted version of the MIPAS retrieval algorithm to be found in von Clarmann et al. (2003) and Funke et al. (2005b). Using 3-D ray tracing, a 2-D retrieval grid, and regularisation with respect to altitude and latitude yields the NO density for half an orbit simultaneously for the altitude range from 60–160 km.

The retrieval of NO number densities from the fit of calculated spectra for the NO gamma bands gives yields NO data, which are consistent with previous measurements from Barth et al. (2003) and Minschwaner et al. (2004). A first comparison shows that they are also in reasonable agreement between 60° N and 60° S with the measurements of the MIPAS instrument on board the same satellite for the orbits where both instruments scanned the upper atmosphere (50–150 km). For low geomagnetic activity (low $A_p$) in 2008–2010, we obtain a vertical resolution of 5–10 km in the altitude range 70–150 km and a horizontal resolution of about 9°.

The current approach is successful in retrieving the NO number densities in the mesosphere and lower thermosphere with SCIAMACHY MLT limb scans. The results are in reasonable agreement with previous measurements from other instruments and provide an almost continuous data source for NO in that region from 2008 to 2012. The comparison to the empirical NOEM model shows that the largest deviations occur at times of low solar activity. The consistency among the measurements, however, indicates that the model assumptions, derived from NO observations at times of high solar activity, need further investigation. In the future, latitudinal and longitudinal resolution can be further improved by using two-dimensional detector arrays on new missions. Similarly, such detectors provide longer integration times for the study of weak signals,

Continuous measurements in the MLT region over a long time span are essential to improve our knowledge and understanding of the connection between solar activity and thermospheric NO. This is an important link in the atmospheric system controlling stratospheric ozone and thereby impacting on climate. Observations in the middle atmosphere also allow to follow the NO produced from solar particles in the region around 100 km downwards during the polar winter, in particular during extraordinary atmospheric events such as stratospheric warmings. This facilitates the attribution of anthropogenic influences on the atmospheric composition from the changes resulting from the varying solar activity during a solar cycle. This is of particular value for the validation of climate models with respect to the production of $NO_x$ and its transport.

*Acknowledgements.* S. Bender and M. Sinnhuber thank the Helmholtz-society for funding this project under the grant number VH-NG-624. The IAA team (M. López-Puertas and B. Funke) was supported by the Spanish MINECO under grant AYA2011-23552 and EC FEDER funds. The SCIAMACHY project was funded by German Aerospace (DLR), the Dutch Space Agency, SNO, and the Belgium ministry. ESA funded the Envisat project. The University of Bremen as Principal Investigator has led the scientific support and development of SCIAMACHY and the scientific exploitation of its data products. We acknowledge support by Deutsche Forschungsgemeinschaft and Open Access Publishing Fund of Karlsruhe Institute of Technology.

The service charges for this open access publication have been covered by a Research Centre of the Helmholtz Association.

Edited by: H. Worden